\documentclass[prd,twocolumn]{revtex4}
\usepackage{graphicx}
\usepackage{amsmath}
\usepackage{amssymb}
\usepackage{comment}
\usepackage{color}
\usepackage[svgnames]{xcolor}
\usepackage{caption}

\begin{document}
\title{Performance of Particle Swarm Optimization on the fully-coherent all-sky search for gravitational waves from compact binary coalescences}
\author{Thilina S. Weerathunga}
\affiliation{Dept. of Physics and Astronomy, University of Texas San Antonio, One UTSA Circle, San Antonio, TX 78249}
\author{Soumya D.~Mohanty}
\affiliation{Dept. of Physics and Astronomy, University of Texas Rio Grande Valley, One West University Blvd.,
Brownsville, Texas 78520}

\begin{abstract}
Fully-coherent all-sky search for gravitational wave (GW) signals from the coalescence of compact object binaries is a computationally expensive task.  Approximations,
such as semi-coherent coincidence searches, are currently used to circumvent the computational barrier with a concomitant loss in sensitivity. We explore the effectiveness
of Particle Swarm Optimization (PSO) 
in addressing this problem. Our results, using a simulated network of detectors with initial LIGO design sensitivities and a realistic signal strength, show that PSO can successfully 
deliver a fully-coherent all-sky search 
with $< 1/10$  the number of
likelihood evaluations needed for a grid-based search. 
\end{abstract}
\maketitle

\section{Introduction} 
\label{intro}


The  detection of gravitational waves, announced in 2016 by the LIGO-Virgo Scientific Collaboration\cite{Abbott_Detection_16}, has launched the new era of gravitational wave (GW) astronomy.
The detected signals, known as GW150914~\cite{GW150914_16_1, GW150914_16_2} and GW151226~\cite{GW151226_16_1}, are best matched by  Compact Binary Coalescence (CBC) sources consisting of the inspiral and merger of black holes. The signals were detected by the two aLIGO~\cite{Abbott_Advanced_LIGO}  detectors, which are currently undergoing 
commissioning to reach their design sensitivity.
Over the next few years, aLIGO will be joined by a worldwide network of comparable sensitivity detectors, namely, advanced Virgo~\cite{Advanced_Virgo}, KAGRA~\cite{Kagra_Japan} and LIGO-India~\cite{LIGO_India}. Combining the data from this network of geographically 
distributed second generation 
detectors will lead to a better overall sensitivity for CBC signals along with better localization on the sky~\cite{Fairhust_11}. Prompt localization will enable the study of such events using multiple messengers of information~\cite{Abbott_bband_followup_16}.

Given that theoretically computed waveforms for CBC signals are sufficiently reliable over a broad parameter range~\cite{Sathyaprakash_91}, it is natural to use the Generalized Likelihood Ratio Test (GLRT) and 
 Maximum-Likelihood Estimation (MLE)~\cite{Kay_1} for the detection and estimation, respectively,
 of such signals. However, both of these methods involve a
 computationally expensive non-linear and non-convex numerical optimization problem. 
 Applied to the data from a network of detectors, the MLE/GLRT approach -- called a fully-coherent search -- requires the localization of the global maximum of the likelihood over an at least nine dimensional parameter space \cite{Pai_01}, where the computation of the likelihood at each point requires correlations between pairs of time series involving $\sim O(10^4)$ (for initial LIGO) to $O(10^6)$ (for aLIGO) samples. A brute force grid-based search for the global
 maximum is estimated to require $\sim 4\times 10^6$ likelihood evaluations over
 the low component mass range of $1$ to $3$~$M_\odot$ in the case of initial LIGO~\cite{owen_99}, with
 the number becoming substantially higher in the case of aLIGO. (The cost of 
 a grid-based
 search is dominated by the exploration of the low mass range due to longer signal
 durations~\cite{owen_99}.) 
 
The computational bottleneck in the fully-coherent search for 
CBC signals  has restricted the scope of its applicability
so far. Fully-coherent search has either been used only for targeted sky locations~\cite{Harry_11} or
has been approximated by semi-coherent all-sky searches~\cite{bose_thilina}. Semi-coherent searches 
reduce the computational burden by downselecting the number of 
data segments to analyze in a fully-coherent step. The downselection is 
based on
requirements such as the simultaneous crossing of detection thresholds~\cite{Lynch_2015} in at least 
two single-detector (incoherent) searches and closeness of the
estimated signal parameters. As shown in~\cite{Macleod_2016}, a semi-coherent search 
trades-off a significant amount of sensitivity for the 
reduced computational cost, with the detection volume being $\sim 25\%$ smaller
than a fully-coherent search.

The estimates of computational cost above pertain to the case of CBC waveforms for systems
in which the effect of the spins of the binary components are negligible. Moving on to the full
parameter space including spins, a grid-based search simply becomes infeasible due to the exponential dependence of the number of points in a grid on the dimensionality 
of the signal parameter space. 
Current efforts at finding alternatives to grid-based search in this context have focused on Markov Chain Monte Carlo (MCMC) methods~\cite{Gilks_MCMC} and its variants.  
In~\cite{MCMC_2008}, MCMC was applied to the parameter estimation problem for CBC signals when the spin of one binary component is non-negligible. It was found that a typical MCMC chain needs about $O(5\times 10^6)$ iterations (likelihood evaluations) in order to converge reliably. This is again
too 
high a computational cost to allow coverage of all data and MCMC
methods have been used only as a parameter estimation step following a candidate detection.

Fast methods for producing rapid estimates of sky location of a CBC source
have been developed~\cite{Singer_2016, Cornish_2016,Pankow_15}. 
In these methods, the subset of so-called intrinsic 
signal parameters, which are responsible for the bulk of the cost of 
a grid search,  are simply replaced  by the values estimated in the 
computationally cheaper single detector searches. Under this approximation, the cost of
evaluating the likelihood over the remaining parameters 
can be reduced significantly. 
Other studies~\cite{Cannon_10,Scott_11} have 
shown that it is possible to find alternative representations of CBC signals that 
speed up the computation of the likelihood function. 

Particle Swarm Optimization (PSO)\cite{Kennedy_95} is a global optimization method that has proven to be effective across a wide range of 
 application areas~\cite{Shi_01} including astronomy, such as CMBR analysis~\cite{CMBR} 
 and gravitational lensing~\cite{G_Lensing}. 
 In the context of 
 GW data analysis, PSO was first used in~\cite{pso_cbc} 
 for
 the single detector CBC search problem
 where it was shown 
 to be effective in locating the global maximum of the likelihood despite its ruggedness.
 In Pulsar Timing Array (PTA) based GW detection, PSO was
 used successfully in optimizing an extremely rugged likelihood function
 over a 12-dimensional 
 search space~\cite{pso_pta}. Other successful applications of PSO in GW data analysis are growing at a steady rate~\cite{c_galactic_binaries}. 
 
 In this paper, we explore the effectiveness of PSO in a fully-coherent all-sky CBC search.
For the purpose of testing PSO, we take a four-detector network, each having 
the initial LIGO design noise Power Spectral Density (PSD)~\cite{Lazzarini_96}, and use the 2-PN binary 
inspiral waveform. We investigate the effectiveness of the 
GLRT and MLE as found by PSO for detection and parameter estimation 
respectively.

The rest of the paper is organized as follows. Sec~\ref{dataModel} describes the data and signal models used in the paper. Sec~\ref{network} presents the objective function to be optimized in
a fully-coherent all-sky search. 
The PSO algorithm is described in Sec~\ref{pso}. {The simulation setup and results are
described in Sec~\ref{Results}. Our 
conclusions are presented in Sec ~\ref{Conclusion}.

\section{Data and signal models}
\label{dataModel}

In the following, the Fourier transform of a function, $a(t)$, of time is
denoted by $\widetilde{a}(f)$.
The strain time series recorded by the ${\rm I}^{\rm th}$ GW detector in 
a network of $N$ detectors is 
\begin{equation} 
\begin{aligned}
x^{I}(t) = h^{I}(t) + n^{I}(t);
 \\
\end{aligned}
\end{equation}
where $h^I(t)$ is the detector response to the incident GW and $n^I(t)$ denotes detector
noise. We will assume that $n^{I}(t)$ is a realization of a zero-mean, stationary Gaussian
stochastic process,
\begin{eqnarray}
E[n^{I}(t)] &=& 0;\\ 
E[\tilde{n}^{I}(f)\;(\tilde{n}^{I}(f^{'}))^*] &=& \frac{1}{2}S_n(f)\delta(f-f^{'})
\end{eqnarray}
with $S_n(f)$ denoting the one-sided noise power spectral density (PSD). It does not carry a detector
index in this paper because we assume identical PSD for all the detectors.

We use the Earth Centered Earth Fixed Frame  (ECEF)~\cite{Leick_04} to define the geometry of the GW detector network and sources. In terms of the TT-gauge polarization waveforms $h_+(t)$ and $h_\times(t)$, the strain response of the $I^{th}$ detector is given by,
\begin{eqnarray}
h^I(t) &=& F_+^I(\alpha,\delta,\psi) h_+(t-\Delta^I) \nonumber\\
& & + F_\times^I (\alpha, \delta,\psi) h_\times(t - \Delta^I)
\;.
\end{eqnarray}
Here, $\alpha$, $\delta$ are the azimuthal and polar angles that define
the unit vector pointing to the source,
\begin{eqnarray}
\widehat{n}  &=& (\cos\alpha \cos\delta,\; \sin \alpha \cos \delta,\; \sin \delta)\;,
\end{eqnarray}
with the direction of propagation of the GW plane wave being $-\widehat{n}$.
The polarization angle $\psi$ gives the orientation of the  wave frame axes
orthogonal to $\widehat{n}$ with respect to the fiducial basis,
\begin{eqnarray} 
    {e}_x^R &=& (\sin \alpha,\; -\cos \alpha,\; 0)\;,\\
    {e}_y^R &=& (-\cos \alpha \sin \delta,\;-\sin\alpha \sin\delta,\;\cos\delta)\;,
    \label{eq:basis_vector_radiation_frame}
\end{eqnarray}
in the same plane. 
For CBC sources, the minor and major axes of the ellipse formed by the projection of the orbit of the binary on the sky provide the preferred orientation for the wave frame axes.
For generic sources, $\psi$ can be set to zero.
$\Delta^I$
is the time delay between the arrival of the signal at the ECEF origin and the detector,
\begin{eqnarray}
\Delta^I &=& \frac{r^I.\;\widehat{n}}{c} \;,
\end{eqnarray}
where $r^I$ is the position vector of the detector in the ECEF.

$F_+^I$ and $F_\times^I$ are called the antenna pattern functions and they are given by,
\begin{eqnarray}
    \begin{pmatrix}
        F^I_+ \\
        F^I_\times{}
    \end{pmatrix} &=& 
    \begin{pmatrix}
        \cos 2\psi \;\; \sin 2\psi \\
        -\sin 2\psi \;\; \cos 2\psi \\ 
    \end{pmatrix}
    \begin{pmatrix}
        U^I_+\\
         U^I_\times\\
    \end{pmatrix}.
\end{eqnarray}
Here, $U^I_+$ and $U^I_\times$ depend on $(\alpha,\delta)$~\cite{Jaranowski_96} and are 
defined by,
\begin{eqnarray}
\begin{aligned}
    U_+^I&=&\overset\leftrightarrow{\epsilon}_+:\overset\leftrightarrow{d^I};
        \quad
     U_\times^I&=&\overset\leftrightarrow{\epsilon}_\times:\overset\leftrightarrow{d^I}; 
\end{aligned}
\end{eqnarray}
where $\overset\leftrightarrow{a}$ denotes a tensor and $:$ denotes the contraction operation on
tensors. 
The polarization basis tensors $\overset\leftrightarrow{\epsilon}_{+,\times}$ are given by,
 \begin{eqnarray} 
    \overset\leftrightarrow{\epsilon}_+ &=& e_x^R\otimes e_x^R -  e_y^R\otimes e_y^R \;,\\
    \overset\leftrightarrow{\epsilon}_\times &=& e_x^R\otimes e_y^R +   e_y^R\otimes  e_x^R\;,
    \label{eq:basis_tensor}
\end{eqnarray}
while
$\overset\leftrightarrow{d^I}$ is the detector tensor 
\begin{eqnarray}
    \overset\leftrightarrow{d^I} &=&\frac{1}{2}
    (\widehat{u}^I\otimes\widehat{u}^I - \widehat{v}^I\otimes\widehat{v}^I)\;,
    \label{eq:detector_tensor}
\end{eqnarray}
 where  ($\widehat{u}^I$, $\widehat{v}^I$) are unit vectors
 along the arms of the detector. 
 
 The relation between 
 the responses of all the detectors in a network, with each appropriately
 time-shifted to compensate for 
 the delay $\Delta^I$, to the incoming GW strain signal 
 can be expressed in the following compact form.
 \begin{eqnarray}
 \left(
 \begin{array}{c}
   h^1(t)\\
   h^2(t)\\
   \vdots\\
   h^N(t)
 \end{array}
 \right) & = & {\bf F}\left(
 \begin{array}{c}
   h_+(t)\\
   h_\times(t)
 \end{array}
 \right)\;,
 \label{antenna_pattern_matrix}
 \end{eqnarray}
 where the $I^{\rm th}$ row of ${\bf F}$, called the {\it antenna pattern matrix}, 
 contains $(F_+^I, F_\times^I)$. It is know that
 ${\bf F}$ can become rank-deficient for 
 certain parts of the sky, leading to an ill-posed inverse 
 problem that can have a significant effect ~\cite{Klimenko_05,Mohanty_06,Rakhmanov_06} on 
 parameter estimation errors. The rank-deficiency of ${\bf F}$ is quantified in terms
 of its condition number. 

 In this paper, we use a four-detector network consisting of the two LIGO detectors at Hanford (H) 
 and Livingston (L), Virgo (V) and Kagra (K). We assume the  initial LIGO design PSD~\cite{Lazzarini_96} for the noise in each detector. 
 The orientations and locations
 of the detectors, provided in Table~\ref{tab:unit_v},  match their real-world values.
\begin{table}
\begin{tabular}{|c|c|c|c|c|}\hline 
	& Hanford & Livingston & Virgo & Kagra \\ \hline
	$u_x$ &-0.2239&-0.9546&-0.7005&-0.4300\\
	$u_y$ &0.7998 &-0.1416&0.2085 &-0.8363\\
	$u_z$ &0.5569 &-0.2622&0.6826 &0.3400\\
	\hline
	$v_x$ &-0.9140&0.2977 &-0.0538& 0.6821\\
	$v_y$ &0.0261 &-0.4879&-0.9691&-0.0542\\
	$v_z$ &-0.4049&-0.8205&0.2408 & 0.7292\\
	\hline
	$x$&-2.1614e+6 &-7.4276e+4 &4.5464e+6&-3.7769e+6 \\
    $y$&-3.8347e+6 &-5.4963e+6 &8.4299e+5 &3.4839e+6 \\
    $z$&4.6004e+6  & 3.2243e+6 &4.3786e+6 &3.7667e+6
    \\
    \hline
\end{tabular}
	\caption{\label{tab:unit_v}Location and orientation vectors for
	the detectors used in this paper. The vector components are specified in the ECEF. The components ($u_x$,$u_y$,$u_z$) and ($v_x$,$v_y$,$v_z$) are for the unit vectors $\widehat{u}^I$ and $\widehat{v}^I$ along the detector arms. The components $(x,y,z)$ of the position vector $r^I$ are in {meters}. The data in this table was obtained from~\cite{detector_info_LIGO, detector_info}.}
\end{table}

Our choice of the initial LIGO PSD allows data realizations to 
be considerably shorter in length than what is needed for the PSDs of
advanced detectors.
This reduces computational costs, which is appropriate for a
first investigation of PSO in the context of a fully coherent search.

\subsection{Restricted 2-PN signal}
The signal model used in this paper is the restricted 2-PN waveform 
from a circularized binary consisting of non-spinning compact objects. The
phase of the signal is calculated up to order {$(v/c)^4$} in the 
post-Newtonian expansion but the amplitude modulation
is calculated only at the lowest (Newtonian) order~\cite{Blanchet_95}.


The GW polarization waveforms can be expressed conveniently 
in the Fourier domain using the stationary phase approximation~\cite{Sathyaprakash_91},
\begin{eqnarray}
    \widetilde{h}_{+}(f) &=& \frac{\mathcal{A}_f}{r}\frac{(1+\cos^2\iota)}{2}f^{-7/6} \exp[{-i}\Psi(f)] \\
    \widetilde{h}_{\times}(f) &=& \frac{\mathcal{A}_f}{r} \cos \iota f^{-7/6} \exp[{-i}(\Psi(f)+\pi/2)]
\end{eqnarray}
where r is the distance to the source and the phase $\Psi(f)$ is given by, 
\begin{eqnarray}
\label{2pn_phase}
    \Psi(f) &=&  2\pi f t_c - {\phi_c} - \pi/4 + \sum\limits_{j=0}^4 \alpha_j\bigg(\frac{f}{f_*}\bigg)^{(-5+j)/3} \;.
\end{eqnarray}
The functional form of 
$\alpha_j$ is given in Appendix~\ref{app_A}. They only depend on 
the component masses $m_1$ and $m_2$ through the {\it chirp times} 
$\tau_0$ and $\tau_{1.5}$.
$t_c$ is the time at which the end of 
the inspiral signal arrives at the ECEF origin. 
The amplitude $\mathcal{A}_f$ depends on 
the component
masses $m_1$ and $m_2$. The coalescence phase of the signal is given by $\phi_c$.
It is possible to absorb  $r$, $\phi_c$, $\psi$ and $\iota$  in a 
new set of parameters $A_k$, ($k = 1, \ldots,4$), giving
\begin{eqnarray}
\begin{aligned}
	 {h}^{I}(t) &=& \sum_{k = 1}^{4}{A_k}{h_k}^I(t)\;, 
\end{aligned}
\label{eq:Newform}
\end{eqnarray}
with,
\begin{widetext}
\begin{eqnarray}
A_1 &=& \frac{1}{r}\bigg(\beta\cos \phi_c\cos 2\psi   
- \cos \iota \sin \phi_c \sin 2\psi   \bigg),\quad
A_3 =-\frac{1}{r}\bigg(\beta \sin \phi_c \cos 2\psi  + \cos \iota \cos \phi_c  \sin 2\psi  \bigg),
\\
A_2 &=& \frac{1}{r}\bigg( \beta\cos \phi_c\sin 2\psi  + \cos \iota \sin \phi_c \cos 2\psi   \bigg), \quad
A_4 = -\frac{1}{r}\bigg(\beta \sin  \phi_c \sin 2\psi   - \cos \iota \cos \phi_c \cos 2\psi   \bigg), 
\label{eq:Extr_Amplitudes}
\end{eqnarray}
\end{widetext}
where 
$\beta = (1+\cos^2 \iota)/2$. Here, the waveforms ${h_k}^I$, k\;=\;1,...,4, are defined as 
\begin{eqnarray}
{h_1^I}(t) &=& U_+^I h_c(t-\Delta^I),\quad \nonumber
{h_2^I}(t) = U_\times^I h_c(t-\Delta^I),\\  \nonumber       
{h_3^I}(t) &=& U_+^I h_s(t-\Delta^I),\quad \nonumber
{h_4^I}(t) = U_\times^I h_s(t-\Delta^I),\nonumber
\label{eq:New_Polar_Forms}
\end{eqnarray}
with,
\begin{eqnarray}
\label{eq:temp_cos_f}
\widetilde{h}_c(f) &=& {\mathcal{A}_f}f^{-7/6} \exp[{-i}\Psi(f)|_{\phi_c = 0}],\\ 
\label{eq:temp_sin_f}
\widetilde{h}_s(f) &=& -i\widetilde{h}_c(f).
\end{eqnarray}
\section{Fully-coherent all-sky search}
\label{network}

Under our assumption of Gaussian, stationary noise, the log-likelihood Ratio (LLR)~\cite{Helstrom_95} for the $I^{\rm th}$ 
detector is given by, 
\begin{eqnarray}
    {\rm ln}\; \lambda^{{I}} &=& {\langle}x^{I}|h^{I}{\rangle} - \frac{1}{2}{\langle}h^{I}|h^{I}{\rangle}\;,
\end{eqnarray}
where, 
\begin{eqnarray}
\label{eq.dot_product}
{\langle}\;p\; |\; q\;{\rangle}  &=& 4 \; Re \int_{0}^{\infty} df\; 
\frac{\widetilde{p}(f)\widetilde{q}^\ast(f)} {{S_n}(f)}.
\end{eqnarray}
If we assume the noise in different detectors 
to be statistically independent, the log-likelihood for an $N$ detector network is given by,
\begin{eqnarray}
		\rm {ln}\;\lambda^{(N)} &=& \sum_{I=1}^{N}\left[ {\langle}x^I|h^I{\rangle} - \frac{1}{2}{\langle}h^I|h^I{\rangle}\right]\;.
\label{eq:NetworkLH}
\end{eqnarray} 
Substituting from Eqs.~\ref{eq:Newform},~\ref{eq:Extr_Amplitudes}, and~\ref{eq:New_Polar_Forms}  we get,
\begin{eqnarray}
    \ln \lambda^{(N)} &=& \mathbf{A^T H - \frac{1}{2} A^T M  A}
    \label{eq:NetworkLH_Compact}
\end{eqnarray}
where $\mathbf{A}=(A_1, A_2, A_3, A_4)^{\rm T}$, $\mathbf{H}=(H_1, H_2, H_3, H_4)^{\rm T}$, and
$H_a= \sum\limits_{I=1}^N {\langle}x^I|h_a^I{\rangle}$ for $a = 1,\ldots,4$.
The matrix $\mathbf{M}$ is given by
\begin{eqnarray}
\textbf M &=& 
         \begin{pmatrix}
     \cal{A} & \cal{B} &  0 & 0 & \\
     \cal{B} & \cal{C} &  0 & 0 & \\
           0 & 0 &  \cal{A} & \cal{B} & \\
           0 & 0 &  \cal{B} & \cal{C} &\\
           \end{pmatrix}
           \label{eq:Network_Matrix}
\end{eqnarray}
where,
\begin{eqnarray}
   {\cal A} &=& \sum\limits_{i=1}^N U_+^I U_+^I{\langle}{h^I_c}|{h^I_c}{\rangle}  \\
   {\cal B} &=& \sum\limits_{i=1}^N U_+^I U_\times^I {\langle}{h^I_c}|{h^I_c}{\rangle}  \nonumber \\
   {\cal C} &=& \sum\limits_{i=1}^N U_\times^I U^I_\times {\langle}{h^I_c}|{h^I_c}{\rangle}  \nonumber 
   \label{eq:Net_Matr_Elements}
\end{eqnarray}
It follows that, for a given data realization, the  log-likelihood is a function of the 
parameters $\mathbf{A}$ and $\theta = \{\tau_0, \tau_{1.5}, \alpha, \delta, t_c\}$.

The GLRT statistic is  the global maximum of the LLR over
the parameters $\mathbf{A}$ and $\theta$.
Following~\cite{bose_thilina}, it is denoted by the equivalent 
statistic $\rho_{\rm coh}$,
\begin{eqnarray}
\rho_{\rm coh}^2 & = & 2 \max_{\mathbf{A},\theta}\ln \lambda^{(N)}\;,
\end{eqnarray}
called the {\it coherent search statistic}.
The MLE, $\widehat{\mathbf{A}}$ and $\widehat{\theta}$, of  $\mathbf{A}$ and $\theta$ respectively are the  maximizers.

The
Maximization of the 
log-likelihood can be carried out as,
\begin{eqnarray}
\rho_{\rm coh}^2 & = &  \max_{\theta}\gamma^2(\theta)\;,\\
\gamma^2(\theta) & = & 2 \max_{\mathbf{A}} \ln \lambda^{(N)}\;.
\end{eqnarray}
The inner maximization over $\mathbf{A}$ can be performed analytically, giving
\begin{eqnarray}
   \mathbf{\widehat{A}} &=& \mathbf{M^{-1}H} \; ,
    \label{eq:A_Hat}
\end{eqnarray}
as the solution. Then,
 \begin{eqnarray}
   \gamma^2(\theta) &=& \mathbf{H^T M^{-1} H}.
 \end{eqnarray}
 The outer maximization  over  $\theta$ must be carried out numerically.
For fixed $\{\tau_0, \tau_{1.5}, \alpha, \delta\}$,
the maximization over  $t_c$, 
 can be carried out very efficiently using the 
Fast Fourier Transform (FFT) since $\langle p | q(t-t_c)\rangle$ is a correlation 
operation~\cite{bose_thilina}.
We call 
\begin{eqnarray}
\Gamma^2(\Theta) & = & \max_{t_c} \gamma^2(\theta)\;\Rightarrow\;\rho^2_{\rm coh}=\max_\Theta\Gamma^2(\Theta),
\label{eq:Max_LLH_Analytical}
\end{eqnarray}
with $\Theta = \{\tau_0, \tau_{1.5}, \alpha, \delta\}$, the {{\it coherent fitness function.}}
Its maximization over $\Theta$
 is the main challenge in the implementation of a fully
coherent search. In this paper, we use PSO, described next, to carry out this task.

\section{Particle Swarm Optimization}
\label{pso}

PSO is an optimization method derived from a simplified 
mathematical model of the swarming behavior observed in nature 
across many species. It uses a fixed number of samples (called {\em particles}) of the
function to be optimized (called the {\em fitness function}). The particles are 
iteratively moved in the search space following a set of dynamical
equations. 

\subsection{PSO algorithm}
\label{pso_algorithm}
To provide a rigorous description of PSO, we adopt the following notation
in this section.
\begin{itemize}
    \item $f(x)$: the scalar  fitness function to be optimized, with 
     $x = (x^1, x^2, \ldots, x^d)\in \mathbb{R}^d$. 
    In our case, $x = \Theta$, $f(x)$ is the coherent fitness 
    function $\Gamma^2(\Theta)$ (c.f., Eq.~\ref{eq:Max_LLH_Analytical})  and $d=4$.
    \item $\mathcal{S}\subset \mathbb{R}^d$: the search space
     defined by the hypercube $a^i\leq x^i \leq b^i$, $i = 1, 2, \ldots, d$.
    \item $N_p$: the number of particles in the swarm.
    \item $x_i[k]$: the position of the $i^{\rm th}$ particle
    at the $k^{\rm th}$ iteration.
    \item $p_i[k]$: the best location found by the $i^{\rm th}$ particle over all iterations up to and including the $k^{\rm th}$.
    \begin{eqnarray}
        f(p_i[k]) & = &\max_{j\leq k} f(x_i[j])\;.
    \end{eqnarray}
    \item $p_g[k]$: the best location found by the swarm over all iterations up to
    and including the $k^{\rm th}$.
    \begin{eqnarray}
        f(p_g[k]) & = & \max_{1\leq j \leq N_p} f(x_j[k])
    \end{eqnarray}
\end{itemize}
The PSO dynamical equations are as follows.
\begin{eqnarray}
v_i[k+1] & = & w[k] v_i[k] + c_1 {\bf r}_1 (p_i[k] - x_i[k]) +\nonumber\\
            &&   c_2 {\bf r}_2 (p_g[k] - x_i[k])\;.
            \label{velocityEqn}\\
x_i[k+1] & = & x_i[k] + z_i[k+1]\;,\\
z^j_i[k] & = & \left\{
\begin{array}{cc}
  v_i^j[k] ,  &  -v_{\rm max}^j \leq v_i^j[k] \leq v_{\rm max}^j\\
   -v_{\rm max}^j ,  & 
   v_i^j[k] < -v_{\rm max}^j\\
   v_{\rm max}^j & v_i^j[k] > v_{\rm max}^j
\end{array}
\right.\;,
\end{eqnarray}
Here, $v_i[k]$ is called the ``velocity" of the $i^{\rm th}$ particle, $w[k]$ is a deterministic function known as the inertia weight (see below), $c_1$ and $c_2$
are constants, and ${\bf r}_i$ is a diagonal matrix with 
independent, identically distributed random variables having a uniform distribution 
over $[0,1]$. The second and third terms on the RHS of Eq.~\ref{velocityEqn} are 
called the {\it cognitive} and {\it social} terms respectively. 

The iterations are initialized at $k=1$ 
by independently drawing (i)  $x_i^j[1]$ 
from a uniform distribution over $[a^j,b^j]$, and (ii) 
$v_i^j[1]$ from a uniform distribution over $[-v_{\rm max}^j, v_{\rm max}^j]$.
          
The behavior of particles must be prescribed when they exit the
search space. We adopt the standard ``let them fly" boundary condition under which a particle outside the search space is 
assigned a fitness values of $-\infty$. Since both $p_i[k]$ and 
$p_g[k]$ are always within the search space, such a particle is
eventually dragged back into the search space by the 
cognitive and social terms. The stopping criterion 
adopted is simply that a fixed number, ${N_{\rm iter}}$, 
of iterations are completed.

\subsection{Assessing convergence}
\label{pso_converge_test}
The conditions for a stochastic optimizer  to converge to the global optimum~\cite{Solis_Wets_1981} require that
(a) every measurable subset of the search space be visited at least once, and 
(ii) the fitness value at any iteration is equal or better than the value at
the previous one. 
There are
no stochastic optimization methods, including PSO, that pass  these conditions in a finite number of iterations. 
Hence, convergence to the global maximum  is not 
guaranteed. Note that this does not 
mean that PSO cannot
converge to the global maximum. It simply means that we can never be sure if it has
found the global maximum or not. One can only talk about the 
probability of convergence in a given run of PSO. A simple way to increase
this probability to near certainty is to do a sufficient number of 
independent runs of PSO. If the probability of convergence in a single run is
$P_{\rm conv}$, then the probability of not converging in any of $N_{\rm runs}$ independent runs is $(1-P_{\rm conv})^{N_{\rm runs}}$. 

Another issue with using a stochastic optimization method like PSO is that it is not
directly possible to verify that the end result of PSO is actually close 
to the global maximum or not. This is because verification can only be done using a 
grid search and, for the problem of interest, a grid search may simply be infeasible. 
(Theoretical studies of PSO use benchmark fitness functions for which the location 
of the global optimum is known by construction.)
However, as pointed out in~\cite{pso_cbc}, there exists an indirect way to check if PSO 
is performing satisfactorily in the case
of likelihood maximization with data containing a signal injected with 
known parameters. 
 Since
any parameter estimation method incurs an estimation error caused by the shift of the 
global maximum away from the true signal parameters, the global maximum of the fitness 
must be higher than the fitness value at the true signal parameters, the 
latter being known for simulated data. Thus, we 
can find out if PSO is doing a satisfactory job or not
by checking that it yields
a fitness value greater than the one at the true signal parameters. 

\subsection{PSO tuning}

Stochastic optimizers such as PSO need to trade off wide-ranging {\em 
exploration}
of the search space against {\em exploitation} of a good candidate location.
These two phases are in conflict with each other,
requiring a proper balance in the relative time spent in each phase.
In general, more exploration leads to higher 
computational cost while making it too short leads to premature
convergence to a local maximum. 
Fig.~\ref{psofitness} shows the global
best fitness value evolution for the coherent network analysis problem.
One can see how PSO initially converges rapidly during the 
exploration phase and then slows down while it searches for the best value in a small region during the exploitation phase.
\begin{figure}[]
\includegraphics[scale=0.05]{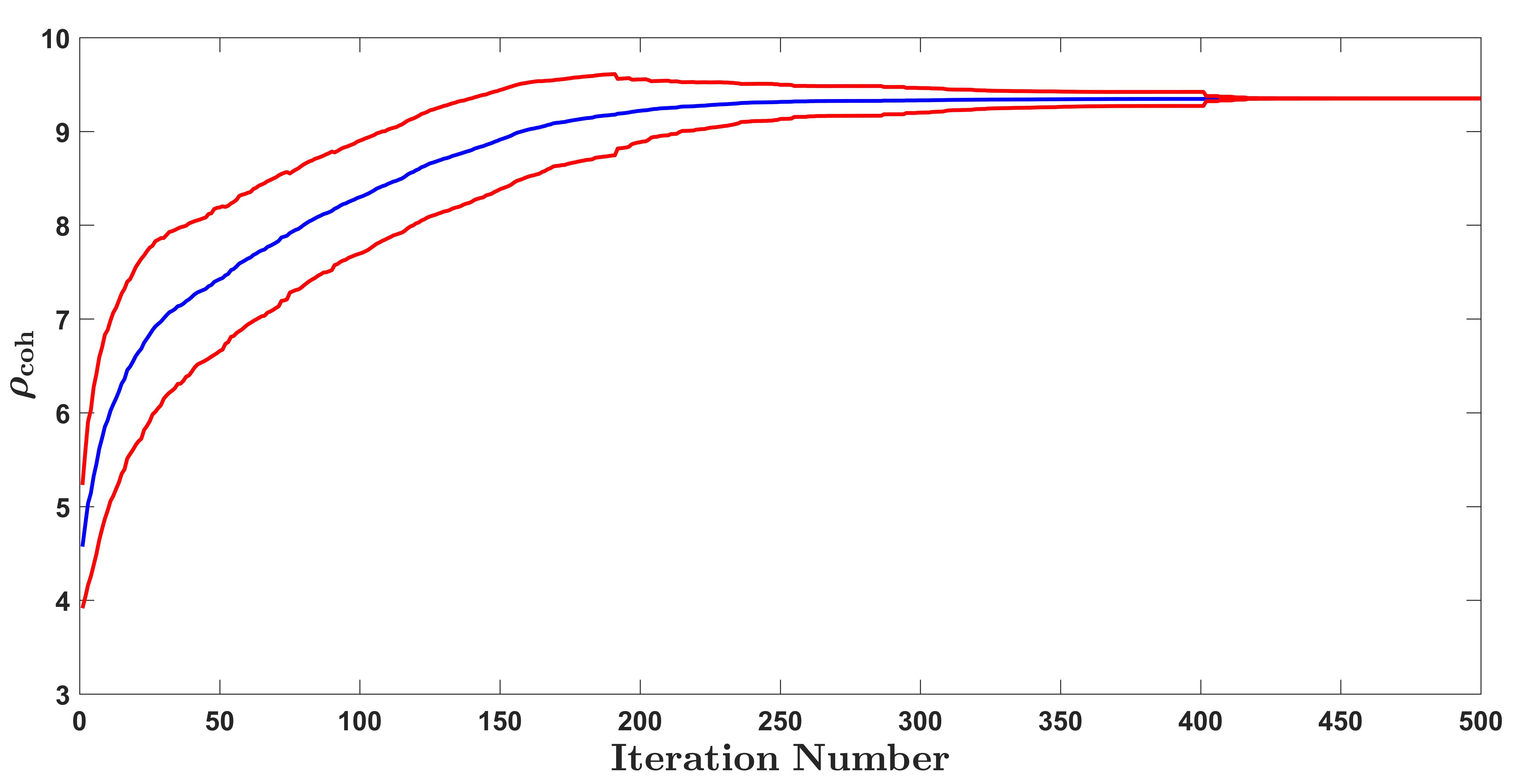}
\caption{Evolution of the mean and standard deviation of the coherent search statistic
for a single data realization. The blue line is the average over
225 independent PSO runs, while the red curves show the $1\sigma$ standard deviation. 
The data realization contains a signal with a 
{ coherent network SNR (defined in Eq.~\ref{coh_snr})
of $9.0$.}
}
\label{psofitness}
\end{figure}

In the version of PSO
described here, the main parameter 
controlling the transition from exploration to exploitation is the inertia weight.
In this paper, the inertia weight
is chosen to decay linearly from a value 
$w_{\rm max}$ at $k = 1$ to $w_{\rm min}$ at $k = {N_{\rm iter}}$. 

An attractive feature of PSO is the apparent robustness of
its parameter values across a wide range of optimization 
problems~\cite{bratton_kennedy}. This greatly reduces the 
effort needed to tune the algorithm for satisfactory 
performance.
We find that, in the optimization problem considered here, 
fairly standard settings~\cite{Poli_07} for the PSO parameters work well. 
The values used for these parameters are
$N_p = 40$, 
$c_1 = c_2 =  2.0$,
$w_{\rm max} = 0.9$, and
$w_{\rm min} = 0.3$.


The only parameter above that required any kind of 
tuning was ${N_{\rm iter}}$. To perform the tuning, we examined the evolution of the the global 
best fitness $f(p_g[k])$ as a function of $k$.  The tuning process starts by picking an ${N_{\rm iter}}$ value 
that is sufficiently deep in the exploitation phase based on a curve such as
Fig.~\ref{psofitness}. We then do 12 independent runs of PSO with this value of ${N_{\rm iter}}$
and find the fraction of runs in which the final global best fitness exceeds the 
fitness at the true signal parameters. This gives an estimate of $P_{\rm conv}$,
the probability of successful convergence. We increase $N_{\rm iter}$ until
$P_{\rm conv} \simeq 0.3$, which gives a probability of failure in 12 independent PSO 
runs of $0.0138$. However, since $P_{\rm conv}$ is estimated using only 12 trials, it is not very accurate. The actual probability of successful convergence is discussed in Sec.~\ref{Results}. Based on this tuning procedure, we set $N_{\rm iter} = 500$.


It is important to note that the PSO algorithm presented
here is considered to be one of the most basic among the 
general class of algorithms that have been proposed under the PSO
meta-heuristic~\cite{Englbrecht_book}. An important variant, for 
example, is the use of a neighborhood best location~\cite{Kennedy_06}
instead of the global best $p_g[k]$. Another variant~\cite{Shi_01} applies a constriction
factor to the equation for $v_i[k]$ instead of clamping its 
components to
the interval $[-v_{\rm max}, v_{\rm max}]$. We did not 
find it necessary to explore these other variants because
the basic version of PSO appears to do a satisfactory job.

\section{Results}
\label{Results}
We test the performance of PSO using simulated realizations of data for the HLVK network
described in Sec.~\ref{dataModel}.  
For each data realization, $N_{\rm runs} = 12$ independent PSO runs are carried out.
The result for each data realization corresponds to the 
output from the run that achieves the best final value of 
the coherent search statistic $\rho_{\rm coh}$. 
The independent PSO runs are executed in parallel, and the choice of $N_{\rm runs}=12$
arises from the 12 processing cores per node in
the  computing cluster that was used for the analysis. 



{The simulated signals are normalized to have a specified
{\it coherent network SNR} (${\rm SNR}_{\rm coh}$), defined as,}
\begin{eqnarray}
\label{coh_snr}
{\rm SNR}_{{\rm coh}}=\gamma(\theta_{\rm true})
=\left[H^T M^{-1} H|_{\hat{\bf A},\theta_{\rm true}}\right]^{1/2},
\end{eqnarray}
{where $\theta_{\rm true}$ denotes the values of $\{\tau_0,\tau_{1.5},\alpha,\delta,t_c\}$ associated with the signal to be normalized.} ${\rm SNR}_{{\rm coh}}$ is related to the optimal network signal to noise ratio (${\rm SNR}_{{\rm opt}}$) by, ${\rm SNR}_{{\rm opt}} \simeq \sqrt{2}\,{\rm SNR}_{{\rm coh}}$
where  ${\rm SNR}_{{\rm opt}}$ is defined as,}
\begin{eqnarray}
    {\rm SNR}_{{\rm opt}} &=&
    {\frac{E[\ln \lambda^{(N)} | H_1]- E[\ln \lambda^{(N)}|H_0]}{
    \left[E[(\ln \lambda^{(N)}-E[\ln\lambda^{(N)}|H_0])^2|H_0]\right]^{1/2}}}\;\\
    & = & 
    {\left[\sum_{I=1}^N \langle h^I | h^I \rangle \right]^{1/2}}\;.
    \label{optimalSNR}
\end{eqnarray}
Here, $E[X | A]$ denotes the conditional 
expectation of a random variable  $X$ given condition $A$.  $H_0$ and $H_1$ correspond to the cases where
a signal is, respectively, absent or present in the data. 
{Normalization using 
${\rm SNR}_{\rm opt}$ assumes the best-case scenario where all the signal 
parameters, including ${\bf A}$, are known {\em a priori}, while normalization 
with ${\rm SNR}_{\rm coh}$ relaxes this unrealistic assumption somehwhat.}
 
 We pick several 
 combinations of binary component masses and source sky locations to generate data realizations containing
 signals. We label these combinations
 using the scheme M$a$L$b$, where $a\in \{1,2\}$ and $b\in \{1,2,3,4,5,6\}$. 
 M1 and M2 refer to the pair of binary component masses $(1.4 M_\odot, 1.4 M_\odot)$
 and $(4.6 M_\odot, 1.4 M_\odot)$ respectively. L$b$ refers to the source sky 
 location, for which  the values used are listed in 
 Table~\ref{tab:table_simulation}. 
 
 For each set, M$a$L$b$, of parameters, 120
 data realizations are generated.
 In all cases, the signals are normalized to have {${\rm SNR_{{\rm coh}}} = 9.0$ (equivalent to ${\rm SNR}_{\rm opt}=12.7$). }
 In all cases, the signals
 correspond to face-on binaries with $\iota = 0 $. As such, $\psi$ gets absorbed 
 into the coalescence phase parameter which is set to
 $\phi_c= \pi/3$ radians.

 \begin{table}
\caption{\label{tab:table_simulation} Source sky locations used in the simulations.
The first column lists the label assigned to the location.
The second and third columns list the azimuthal and polar angles of the source location
in the ECEF. 
{The condition number of the antenna pattern matrix at each
chosen source location is listed in the last column.} }
\begin{tabular}{|c|r|r|c|}\hline 
Label &
$\alpha$ (deg) & $\delta$ (deg) 
&  $\rm log_{10}(Condition\;\; number)$
\\
\hline
L1 & 80.79 & -29.22 & 0.9711
\\
L2 & -128.34 & -33.80  & 0.6079
\\
L3 & -81.93 &  13.75  & 0.6008
\\
L4 & 32.09 & -53.86  & 3.2436
\\
L5 & 150.11 & -60.16  & 0.0066
\\
L6 & -122.61 &  41.25  & 1.6159\\
\hline
\end{tabular}
\end{table}
The degree of ill-posedness in the inverse problem of coherent network analysis can be measured in terms of the condition number~\cite{ Mohanty_06,Klimenko_05,Rakhmanov_06} of the antenna pattern matrix {\bf F}. Since
the antenna pattern functions depend on the sky location of a source, so does the condition number. The sky locations chosen in Table~\ref{tab:table_simulation} sample the 
 condition number across a range of values, corresponding to a well conditioned (low) to poorly conditioned (high) inverse problem. 

 For each realization, the data is generated directly in the Fourier domain with a frequency spacing of {0.0156 Hz between consecutive bins and a maximum (Nyquist) frequency of 1024 Hz. In the time
domain, this corresponds to a data segment duration of 64 sec sampled at 2048 Hz. }
The chirp times corresponding to the two sets of masses used in the simulation are
$(\tau_0,\tau_{1.5}) = (24.850,0.866)$~sec and $(9.751,0.728)$~sec respectively. It is 
assumed that the signals are not visible at frequencies below  
{$40$~Hz} due to the steep rise in seismic noise 
below this frequency. Hence, the signal waveform samples are set to zero
 below this frequency. Similarly, 
any inspiral signal terminates when the binary components reach the last stable circular
orbit. We use $700$~Hz, the frequency corresponding to 
the lower total mass system, as the cutoff frequency for all signal waveforms. The 
last stable orbit frequency decreases for higher mass systems but, given the small number
of cycles at high frequencies, it makes no practical difference to the results if a 
uniform cutoff frequency is used.

 For PSO, we use a search range of (i)
 $[0, 43.538]$~sec for $\tau_0$, (ii)$ [0, 1.084]$~sec for $\tau_{1.5}$, (iii) $[-180, 180]$ degrees
 for right ascension $(\alpha)$ and (iii)$[-90, 90]$ degrees
 for declination $(\delta)$.

\subsection{Detection performance}
\label{detection}
It is important to note that the value of the coherent search statistic, $\rho_{\rm coh}$, found by PSO need
not be the actual value, namely, the true 
global maximum of the log-likelihood ratio. Hence, it is important to verify that
$\rho_{\rm coh}$ as found by PSO performs well in terms of detection.

\begin{figure*}	
\includegraphics[scale=0.1]{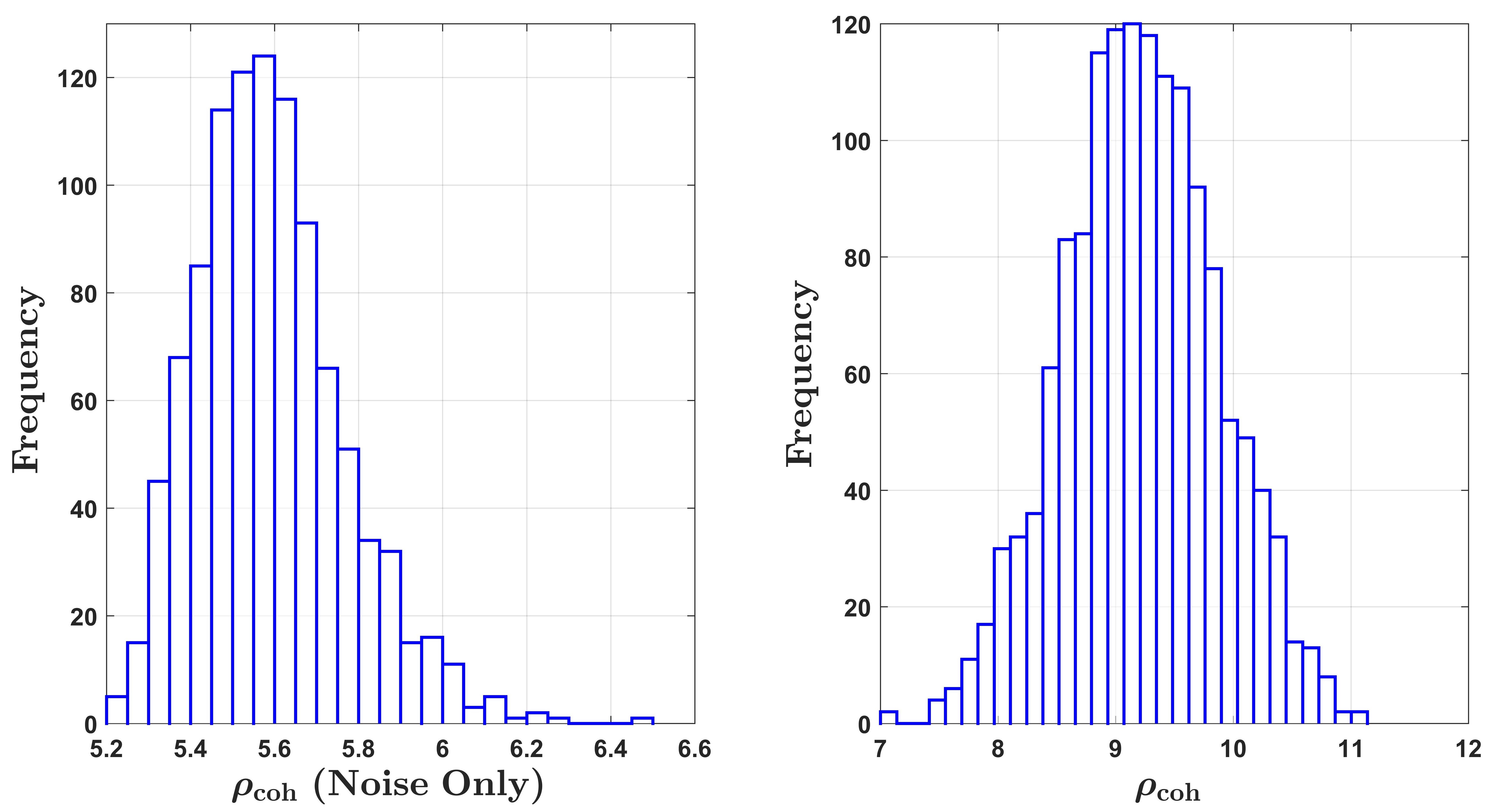}
\captionsetup{justification=centering}
\caption{
\label{fig:Esti_SNR}
{Distribution of coherent search statistic, $\rho_{\rm coh}$, values found by PSO. 
\textbf{Left:} Noise-only case obtained from 1024 independent data realizations. \textbf{Right:} Signal plus noise for {$\rm SNR_{\rm coh} = 9$.}}} For the latter, we have pooled together the values of $\rho_{\rm coh}$ for all the 12 source parameter value sets used in the simulations. This leads to 1440 independent trial values of $\rho_{\rm coh}$ for the histogram on the right.
\end{figure*}

Fig.~\ref{fig:Esti_SNR} shows the distribution of $\rho_{\rm coh}$ found by PSO for data realizations corresponding to 
(i) the null hypothesis ($H_0$): signal absent, and (ii) the 
alternative hypothesis ($H_1$): signal present. 
For the latter,
we have combined the $\rho_{\rm coh}$ values for the 12  source parameter values, M$a$L$b$, used in the generation of data realizations containing signals.
For the number of data realizations used in our simulations, there is no
overlap between the distribution of $\rho_{\rm coh}$ for $H_0$ and $H_1$. 
This suggests that $\rho_{\rm coh}$ found by PSO performs quite well
as a detection statistic and
that it is an acceptable surrogate for the true coherent search statistic.  

\begin{figure}
\includegraphics[scale=0.095]{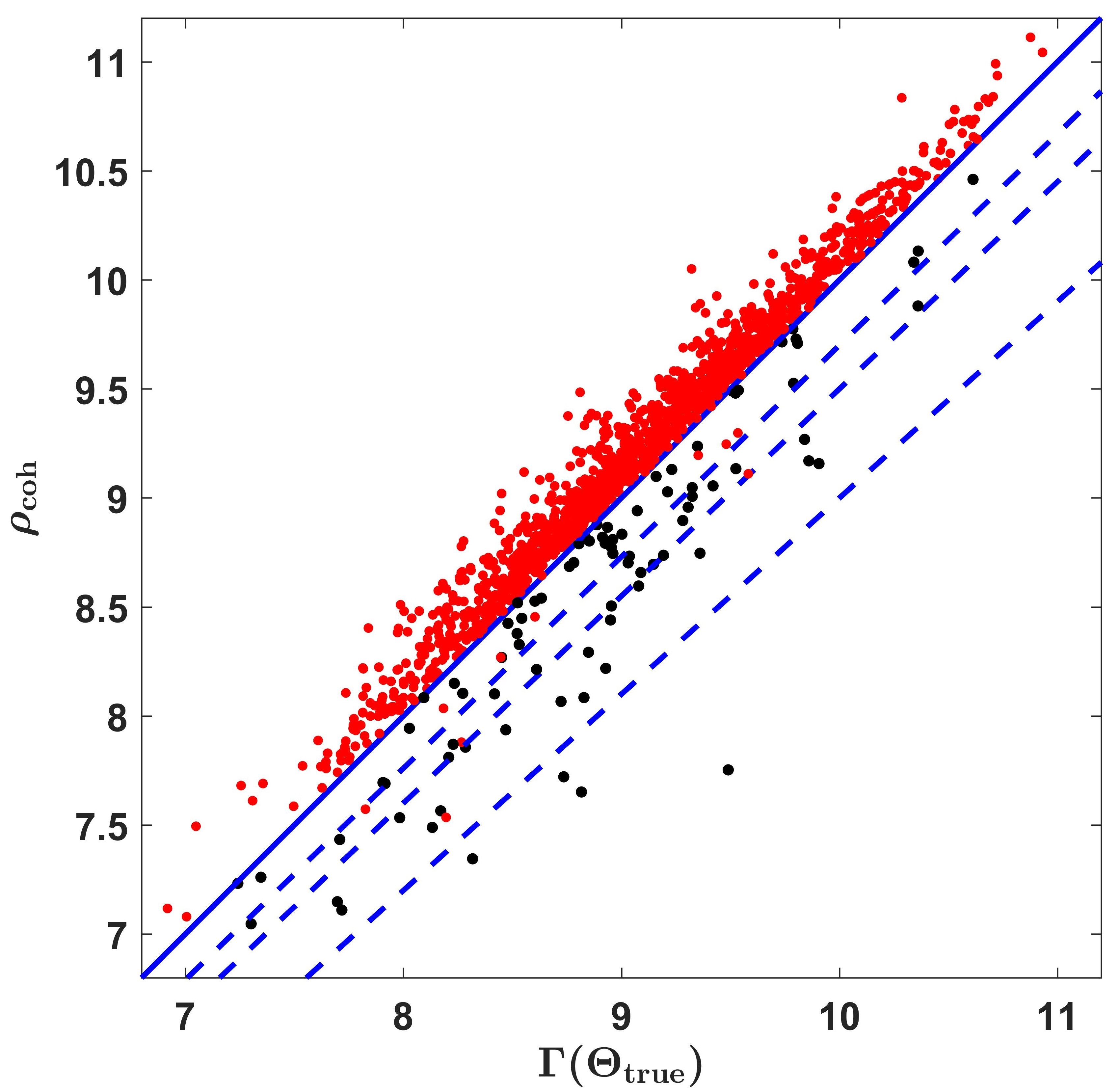}
\caption{
\label{pso_vs_true}Comparsion of the coherent search statistic $\rho_{\rm coh}$ found by PSO
with the coherent fitness value $\Gamma(\Theta_{\rm true})$ at the true signal parameters,
$\Theta_{\rm true}$. Each dot corresponds to one data realization, from a 
total of {1440 } realizations across
all the source parameters used. 
Dashed lines show the $3\%$,$5\%$, and $10\%$ drop from the coherent
fitness value. Black dots indicate data realizations for which $\rho_{\rm coh} < \Gamma(\Theta_{\rm true})$ with $N_{\rm runs} = 12$ indpendent PSO runs, but recovered to 
$\rho_{\rm coh} \geq \Gamma(\Theta_{\rm true})$ when $N_{\rm runs} = 24$. The total 
number of points below the diagonal is 95.}
\end{figure}

Fig.~\ref{pso_vs_true} shows a scatterplot of $\rho_{\rm coh}$ found by PSO against the 
coherent fitness function at the true signal location, $\Gamma(\Theta_{\rm true})$, where 
$\Theta_{\rm true}$ denotes the known parameters of the signal injected into the 
data realization. As discussed 
in Sec.~\ref{pso_converge_test}, $\rho_{\rm coh} \geq \Gamma(\Theta_{\rm true})$ 
indicates that PSO has likely found the global maximum. We find that this condition 
is satisfied in $93.4\%$ of the total number of data realizations when the 
number of independent PSO runs is set to $N_{\rm runs} = 12$. Quantifying the 
departure from this condition in terms of $q = (1 - \rho_{\rm coh}/\Gamma(\Theta_{\rm true}))\times 100$, ${97.0\%}$, ${98.4\%}$, and ${99.7\%}$ of all trials satisfy $q \leq 3\%$, 
$\leq 5\%$, and $\leq 10\%$ respectively. 
When the number of PSO runs is set to $N_{\rm runs} = 24$ for the realizations that
showed a departure from the above condition, the vast majority ended up 
satisfying the condition again, leading to
${99.7 \%}$, ${99.9\%}$, and ${100\%}$ of all trials for $q \leq 3\%$, 
$\leq 5\%$, and $\leq 10\%$ respectively.

While a violation of the condition $\rho_{\rm coh} \geq \Gamma(\theta_{\rm true})$ 
means a failure to locate the global maximum, the drop in $\rho_{\rm coh}$ 
found by PSO from its 
true value may be small enough that the detection threshold is still crossed. This 
would result in the detection of a signal, although it may
not provide a good estimate of its 
parameters. The loss in detection probability can be estimated in terms of the fraction 
of realizations in which $\rho_{\rm coh}$ found by PSO fell below a given
detection threshold while $\Gamma(\Theta_{\rm true})$ stayed above it. Since the 
true $\rho_{\rm coh}$ is always greater than $\Gamma(\Theta_{\rm true})$, the latter condition implies that a detection would have resulted if the true $\rho_{\rm coh}$
were available. For a detection threshold $\eta$, we find that the fractional loss in detection probability, given by
the number of realizations where $\rho_{\rm coh} \leq \eta$ and 
$\Gamma(\Theta_{\rm true}) \geq \eta$  relative to the number where $\Gamma(\Theta_{\rm true}) \geq \eta$,  is between $\simeq 0.9\%$ to 
$\simeq 2.5\%$ for $8.0\leq\eta\leq9.0$.
\subsection{Estimation performance}
\begin{figure*}
\includegraphics[scale=0.185]{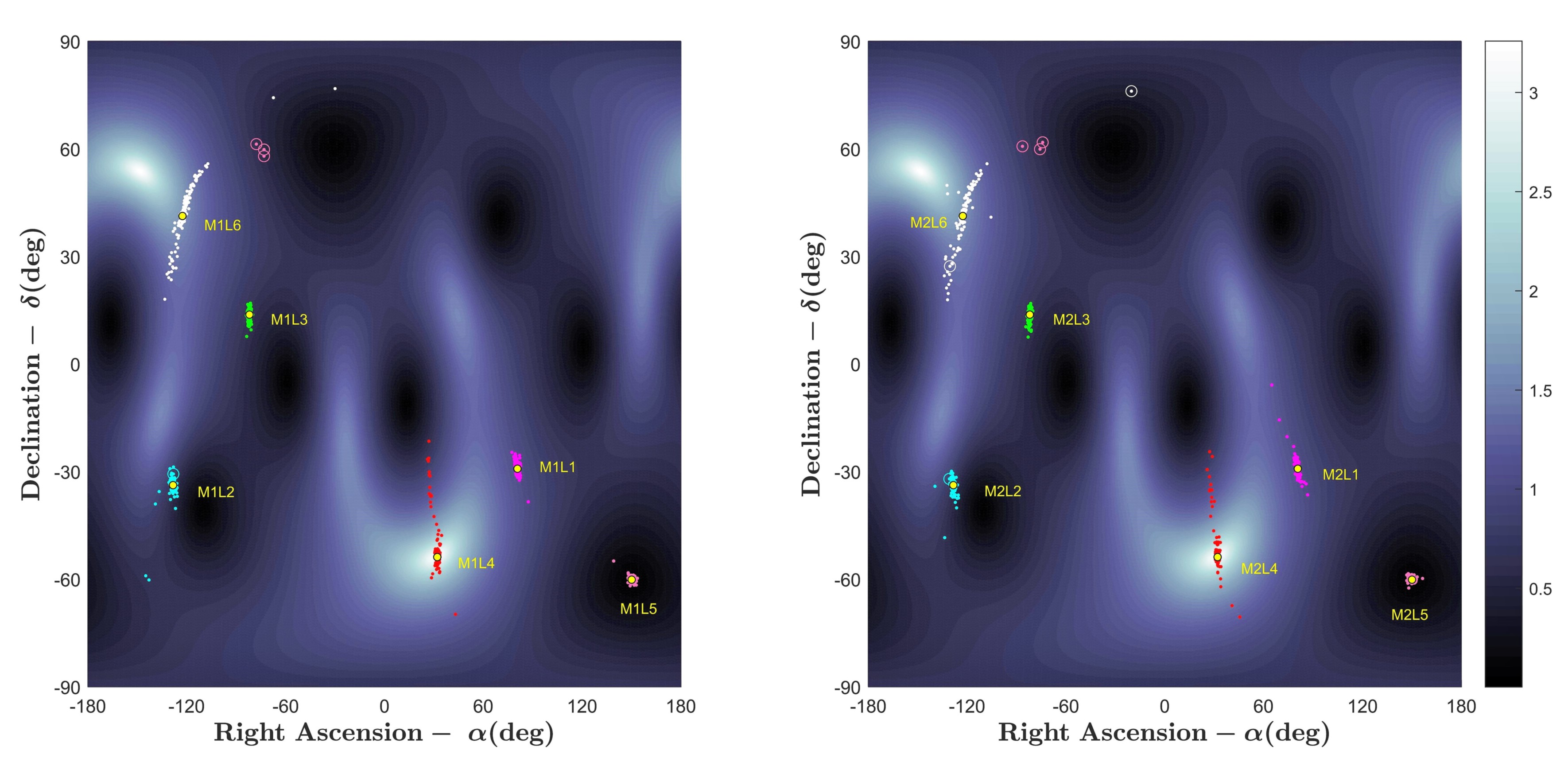}
\caption{\label{fig:Esti_sky}
Estimated sky locations
for the mass sets M1 $(m_1 = 1.4 M_\odot, m_2 = 1.4 M_\odot)$ and M2 $(m_1 = 4.6 M_\odot, m_2 = 1.4 M_\odot)$. The true sky locations are indicated with yellow dots and listed
in Table~\ref{tab:table_simulation}. Estimated sky locations 
associated with a particular true location have  the same color.
Open circles with dots correspond to the 
data realizations where the coherent search statistic found by PSO failed
to exceed the coherent fitness at the true signal parameter,
$\rho < \Gamma(\theta_{\rm true})$, even for $N_{\rm runs} = 24$ independent PSO runs. 
The background gray-scale image in both panels is
identical and shows the condition
number sky map for the HLVK network corresponding to $\psi = \pi/6$. 
}
\end{figure*}

\begin{figure*}	\includegraphics[scale=0.1]{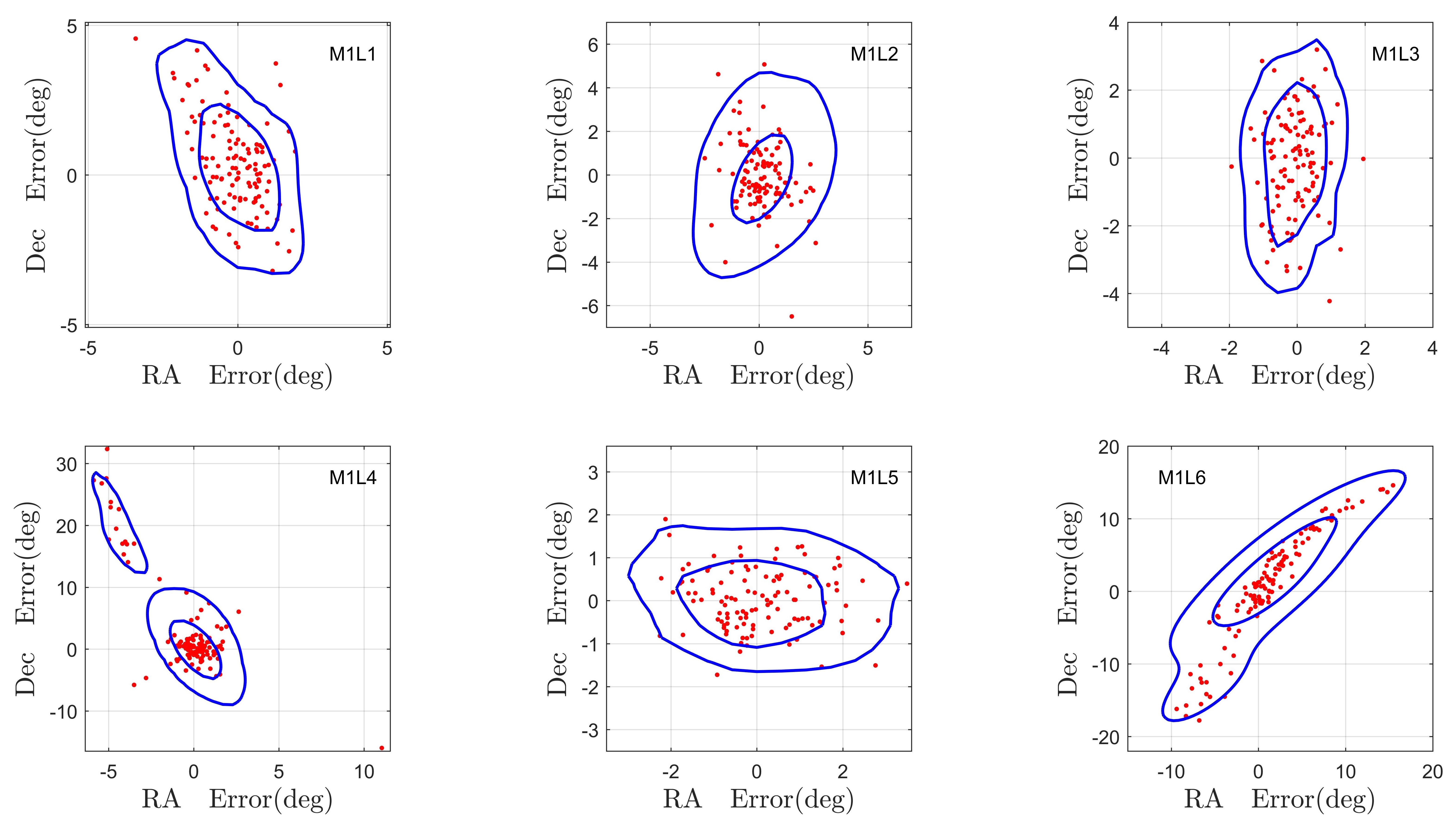}
\caption{\label{fig:KDE_M1}
Estimated sky locations (red dots) associated with
the set M1 $(m_1 = 1.4 M_\odot, m_2 = 1.4 M_\odot)$ of sources. In each panel,
the origin is centered at the true location of the source. The axes show the deviation 
of the estimated values of $\alpha$ and $\delta$ from their true values.
Each panel also shows the contour levels of the bivariate probability density function,
estimated using Kernel Density Estimation (KDE),~\cite{Epanechnikov_67} that enclose $68\%$ and $95\%$ of the points. 
In these figures, the view has been zoomed in to show only the estimated locations
that fall within or around the outer contour.
}
\end{figure*}

\begin{figure*}
\includegraphics[scale=0.1]{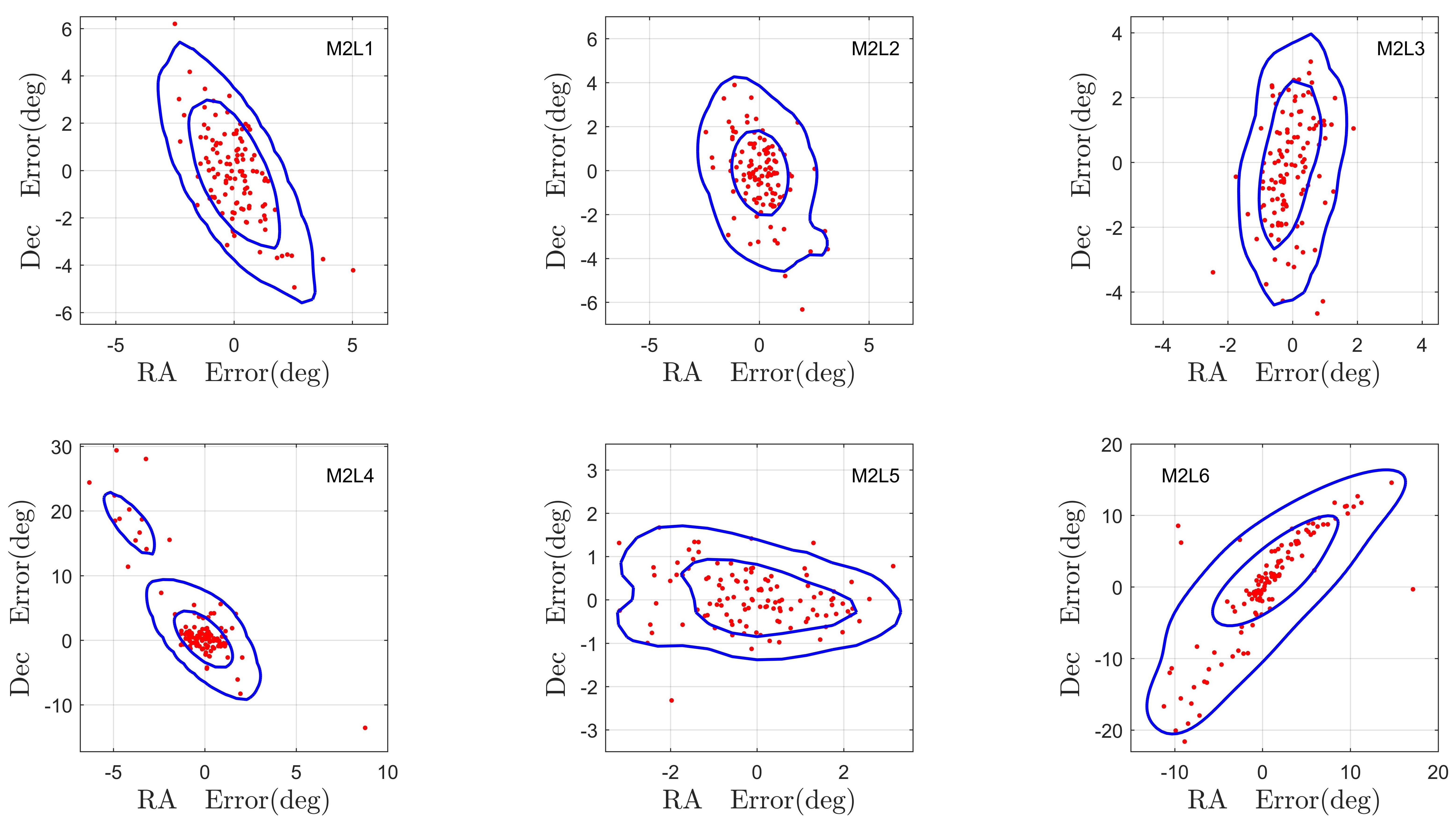}
\caption{\label{fig:KDE_M2}
Estimated sky locations (red dots) associated with
the set M2 $(m_1 = 4.6 M_\odot, m_2 = 1.4 M_\odot)$ of sources. In each panel,
the origin is centered at the true location of the source. The axes show the deviation 
of the estimated values of $\alpha$ and $\delta$ from their true values.
Each panel also shows the contour levels of the bivariate probability density function,
estimated using Kernel Density Estimation (KDE),~\cite{Epanechnikov_67} that enclose $68\%$ and $95\%$ of the points. 
In these figures, the view has been zoomed in to show only the estimated locations
that fall within or around the outer contour.
}
\end{figure*}

The performance of PSO in estimating the sky location of a source is shown in
Fig.~\ref{fig:Esti_sky}, with zoomed in views shown in Figs.~\ref{fig:KDE_M1} and
~\ref{fig:KDE_M2}. It is evident from  Fig.~\ref{fig:Esti_sky} that the 
distribution of sky localization error is strongly influenced by the 
condition number of the antenna pattern matrix (c.f. Eq.~\ref{antenna_pattern_matrix})
at the true location. 
A source location with a low condition number tends to have a
small positional 
error.

We use the area {$16\sigma_\alpha \sigma_\delta \cos\delta$} of the box
of side length {$4\sigma_\alpha$ and $4 \sigma_\delta$}, where $\sigma_\alpha$ and 
$\sigma_\delta$ are the standard deviations in estimates of $\alpha$ and $\delta$ respectively, as a simple measure of position error.
For the 
sources M1L5 and M2L5 that have the lowest condition number, the position errors are 
$10.43$ and $7.32$~${\rm deg}^2$ 
respectively. For the highest condition number location (L4), we get $152.45$ and 
$140.39$~${\rm deg}^2$ for M1 and M2 respectively. The highest position error
occurs not at the highest condition number but at L6, which is the second highest. 
However, our simple measure of position error is wholly inapplicable to these extreme 
locations because the error is distributed along a stretched out region. A proper 
estimation of errors for extreme condition numbers requires a much larger number
of data realizations in order to map out the error region with a sufficient density of sample points. This is postponed to future work pending ongoing work on increasing the 
computational efficiency of our codes. 



\begin{figure*}	\includegraphics[scale=0.2]{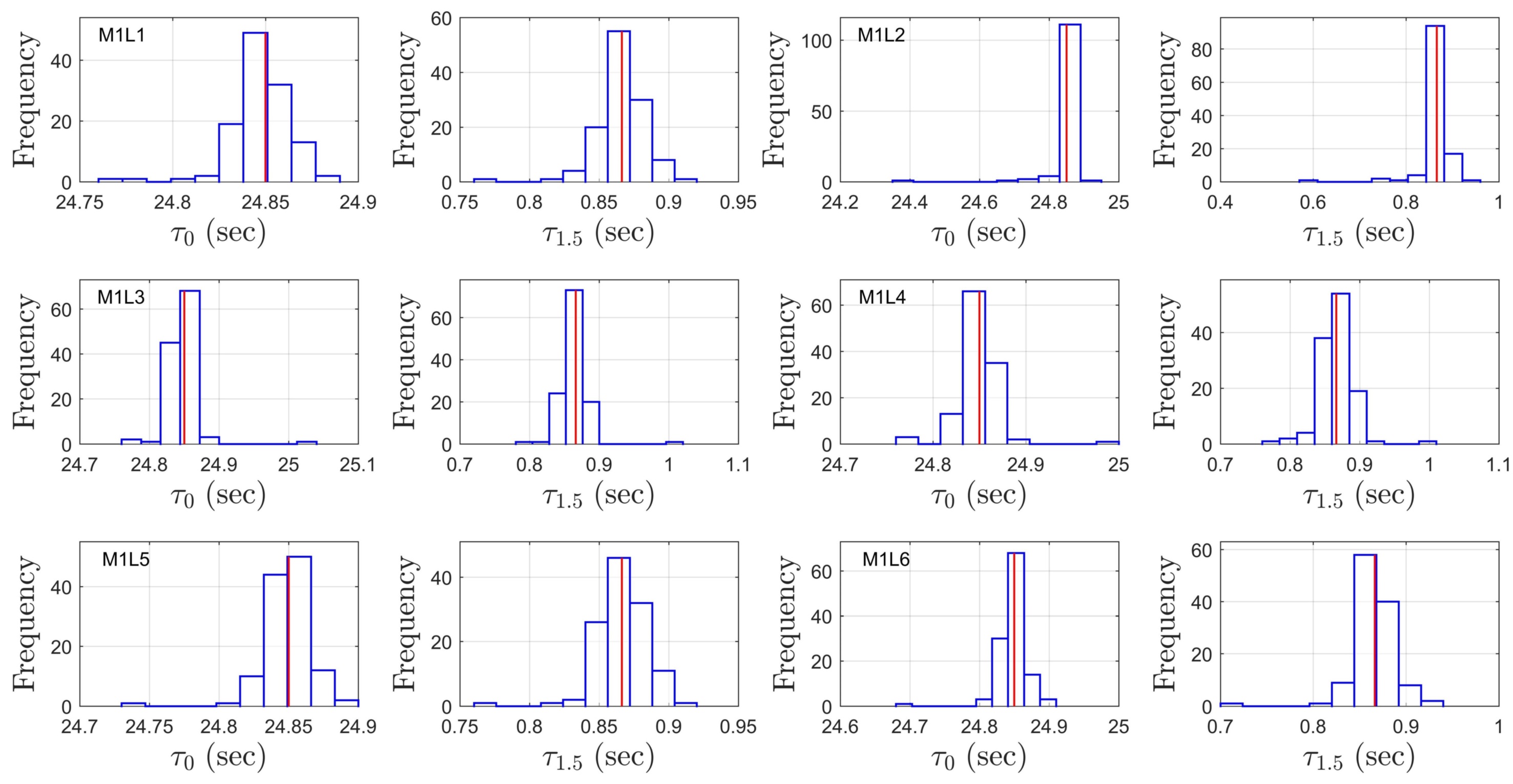}
\caption{\label{fig:M1chirptimes}Histograms of estimated chirp times, $\tau_0$
and $\tau_{1.5}$, for all locations and mass set M1 ($1.4 M_\odot$ and $1.4 M_\odot$).
The true values of the chirp times are shown by the red line in each plot. 
For each source sky location, the $\tau_0$ and $\tau_{1.5}$ distributions are 
adjacent and on the same row, with the $\tau_{1.5}$ distribution always to the right
of the $\tau_0$ one.}
\end{figure*}

\begin{figure*}	\includegraphics[scale=0.22]{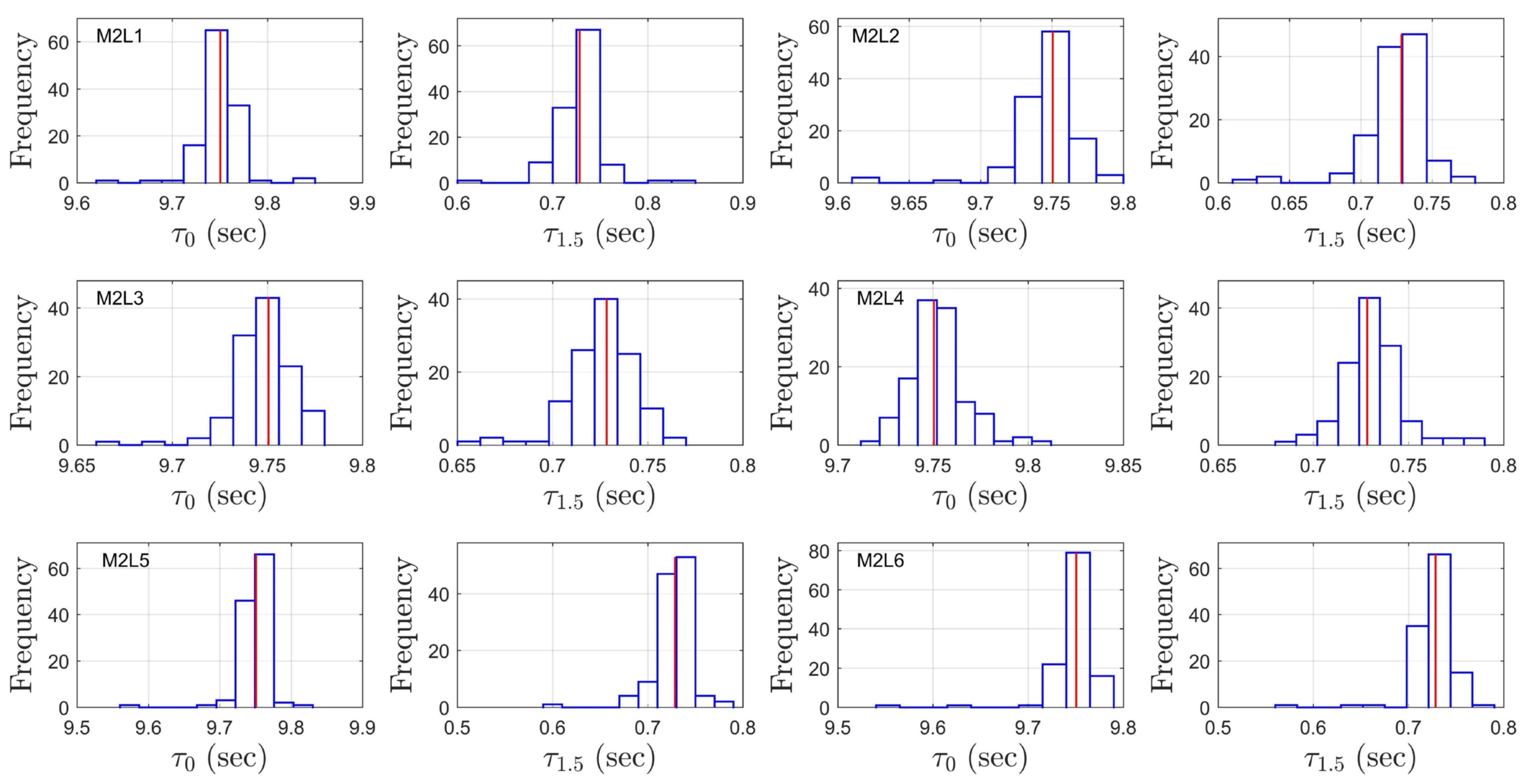}
\caption{\label{fig:M2chirptimes}Histograms of estimated chirp times, $\tau_0$
and $\tau_{1.5}$, for all locations and mass set M2 ($4.6 M_\odot$ and $1.4 M_\odot$).
The true values of the chirp times are shown by the red lines. 
For each source sky location, the $\tau_0$ and $\tau_{1.5}$ distributions are 
adjacent and on the same row, with the $\tau_{1.5}$ distribution always to the right
of the $\tau_0$ one.}
\end{figure*}

Table~\ref{tab:sim_summary} summarizes the marginal distribution of errors up to the second 
moment for all the signal parameters constituting the PSO search space. No clear
trend emerges for the dependence  on condition number of the errors in the chirp time parameters $\tau_0$ and $\tau_{1.5}$. It is likely that resolving a dependence, if any, requires a 
significantly larger number of data realizations. For completeness, the marginal 
distributions are shown in Figs.~\ref{fig:M1chirptimes} and~\ref{fig:M2chirptimes}.
\begin{table*}
 	\caption{\label{tab:sim_summary} Sample mean and standard deviation (SD)
 	of signal 
 	parameter estimates. The estimates of the binary component masses,
 	$m_1$ and $m_2$, are derived from those of the chirp time parameters $\tau_0$ and 
 	$\tau_{1.5}$ using the relations given in~\ref{app_A}. }
 	\begin{ruledtabular}
 		\begin{tabular}{|c|cc|cc|rr|rr|cc|cc|}
 		&\multicolumn{2}{c|}{$\tau_{0}(\rm s)$}
 			&\multicolumn{2}{c|}{$\tau_{1.5}(\rm s)$}
 			&\multicolumn{2}{c|}{$\alpha (\rm deg)$}
 			&\multicolumn{2}{c|}{$\delta (\rm deg)$}
 			&\multicolumn{2}{c|}{$ m_1(M_{\odot})$}
 			&\multicolumn{2}{c|}{$m_2(M_{\odot})$}\\
 			Source &
 			Mean & SD &
 			Mean & SD &
 			Mean & SD &
 			Mean & SD &
 			Mean & SD & 
 			Mean & SD  \\ \hline
 			
 			M1L1 &24.848 & 0.017 & 0.866 & 0.018 & 80.78  & 1.13  & -28.90  & 1.79 & 1.491 &0.138 &1.299&0.089\\
 			
 			M1L2 & 24.846 & 0.048 & 0.863 & 0.036 & -128.61 & 2.57  & -34.40 & 3.68 & 1.502 &0.161 &1.282&0.099\\
 			
  			M1L3 & 24.848 & 0.023 & 0.863 & 0.022 & -82.03 & 0.64  & 13.55  & 1.59 & 1.471 &0.151 &1.311&0.089\\
  			
 			M1L4 & 24.849 & 0.024 & 0.866 & 0.025 & 31.73  & 2.09  & -51.02 & 7.73 & 1.506 &0.174 &1.284&0.104\\
 			
 			M1L5 & 24.849 & 0.018 & 0.866 & 0.018 & 150.09  & 1.56 & -60.10 & 0.84 & 1.499 &0.145 &1.292&0.092\\
 			
 			M1L6 &24.849 & 0.022 & 0.865 & 0.023 & -120.20 & 10.96 & 42.37 & 8.95
 			& 1.498 &0.157 &1.291&0.098\\
 			
 			M2L1  & 9.750 & 0.023 & 0.728 & 0.025 & 80.61   & 2.25  & -29.06 & 3.26 & 4.594 &0.266 &1.406&0.077 \\
 			
 			M2L2  &9.747 & 0.023 & 0.725  & 0.022  & -128.41& 1.47  & -34.17 & 2.09& 4.559 &0.247 &1.415&0.080 \\
 			
 			M2L3 &9.748 & 0.016 & 0.726 & 0.018 & -82.02 & 0.65  & 13.52 & 1.79 & 4.570 &0.189 &1.410&0.052 \\
 			
 			M2L4 &9.752 & 0.015 & 0.731 & 0.016 & 31.83  & 2.15  & -51.74& 6.92 & 4.623 &0.164 &1.396&0.042  \\
 			
 			M2L5& 9.750 & 0.023 & 0.728 & 0.021 & 150.08  & 1.46 & -60.11 & 0.63 & 4.590 &0.239 &1.407 &0.088 \\
 			
 			M2L6 &9.748  & 0.025 & 0.726 & 0.022 & -122.14 & 5.27 & 41.26  & 7.85 & 4.563 &0.272 &1.415 & 0.103
 \end{tabular}
 \end{ruledtabular}
 \end{table*}
Table~\ref{tab:corrcoef} lists the sample correlation coefficients between
pairs of parameters. The sample correlation coefficient between
$\tau_0$ and $\tau_{1.5}$ is $\geq 0.9$ for all the sources considered here.
The strong correlation between chirp time estimates
is well known from studies of single detector searches ~\cite{Sathyaprakash_94}. It is generally assumed from
Fisher information matrix based analyses ~\cite{Singer_16}
that the correlation between the sky angles, $\alpha$ and $\delta$,
and chirp time parameters is negligible. While this result, which strictly 
holds only for asymptotically large {SNRs}, is borne out by our simulation
for the majority of cases, there are some sky locations for both the M1 and M2 sets where this does not hold. 
For example, for M1, there are two locations, L6 and L2, where the sample correlation coefficients are $-0.452$ and
$0.416$ respectively, while it is low (absolute value $\lesssim 0.24$) elsewhere.
Thus, 
the Fisher information may not
be a good predictor of covariances between parameters for all 
source locations. 
\begin{table}
 	\caption{\label{tab:corrcoef} {Sample pair-wise correlation coefficients of 
 	parameters.} The parameter pairs are listed in the headings of the columns. }
 	\begin{ruledtabular}
 		\begin{tabular}{|c|c|r|r|r|r|r|}
 			Source &
 			$(\tau_{0},\tau_{1.5})$ & 
 			$(\alpha,\delta)$ &
 			$(\tau_0,\alpha)$ &
 			$(\tau_0,\delta)$ &
 			$(\tau_{1.5},\alpha)$ &
 			$(\tau_{1.5},\delta)$
 			\\ \hline
 			M1L1 & 0.933 & -0.618 &  0.152& -0.111 & 0.165 & -0.096\\
 			M1L2 & 0.955 &  0.725 &  0.223&  0.435 & 0.105 &  0.310\\
 			M1L3 & 0.959 &  0.174 & -0.231& -0.179 &-0.263 & -0.163\\
 			M1L4 & 0.972 & -0.791 & -0.151&  0.020 &-0.144 &  0.051\\
 			M1L5 & 0.951 & -0.426 & -0.206&  0.120 &-0.229 &  0.150\\
 			M1L6 & 0.926 &  0.806 & -0.452& -0.304 &-0.345 & -0.253\\
 			M2L1 & 0.970 & -0.891 &  0.054& -0.025 & 0.045 & -0.028\\
 			M2L2 & 0.947 & -0.003 &  0.416&  0.171 & 0.259 &  0.222\\
 			M2L3 & 0.935 &  0.305 & -0.067& -0.034 &-0.070 & -0.011\\
 			M2L4 & 0.928 & -0.794 &  0.011&  0.081 & 0.058 &  0.031\\
 			M2L5 & 0.956 & -0.147 &  0.093& -0.122 & 0.024 & -0.095\\
 			M2L6 & 0.959 &  0.821 &  0.018&  0.056 & 0.015 &  0.067\\
 	    \end{tabular}
 \end{ruledtabular}
 \end{table}


\subsection{Computational cost}
Obtaining the coherent fitness value for each PSO particle is the computationally most
expensive step. The calculation of a single fitness value 
requires, (i) the generation of two template waveforms (Eq.~\ref{eq:temp_cos_f}, Eq.~\ref{eq:temp_sin_f}) 
in the Fourier domain, (ii)
taking a sample-wise product of the data with each of the template waveforms (Eq.~\ref{eq.dot_product}), 
and (iii)
taking the inverse FFT of each such product sequence~\cite{bose_thilina}.
The computational cost of each fitness evaluation in the PSO based approach
is identical to those of other 
stochastic optimization algorithms, such as MCMC, that have been 
used for fully-coherent all-sky search.

Among the above operations, the generation of template waveforms is the computationally most expensive step~\cite{Buonanno_09}. 
In situations where  a grid-based 
search for the global maximum is computationally feasible, template waveforms can be 
pre-computed and stored in advance, thus saving the cost of generating waveforms on-the-fly.
Stochastic search algorithms do not use pre-computed waveforms and, hence,
must contend with this extra cost.

Several schemes \cite{Smith_13,Field_14,Purrer_14,Blackman_15}
have been constructed to speed up waveform generation
but we have not implemented any of these in our code so far. 
Besides this, our code is written entirely in Matlab and suffers a penalty in speed
as a result.
Thus, the wall-clock 
execution time of our code is not the correct metric to use for judging the 
computational savings brought about by PSO. The only useful metric for comparing 
PSO with other methods is the total number of 
fitness evaluations. Given the same code for evaluating the fitness 
function, algorithms that get by using a smaller number of evaluations will
automatically have a better execution speed.

Since the termination criteria used for PSO in this paper is simply the number of iterations, there is an upper limit to the total 
number of fitness evaluations of $40 ({\rm number\; of\; particles)} \times 
500 ({\rm number\; of\; iterations}) \times 12 ({\rm number\; of\; PSO\; runs}) = 2.4\times 10^5$. However, the actual 
number of fitness evaluations is generally lower because of the boundary condition used, which allows particles to escape the search space. Particles
outside the search space do not evaluate their fitness until they are pulled
back in. Thus, the number of fitness evaluations can fluctuate across data realizations.
Table~\ref{tab:temp_num} lists a statistical summary of the number of fitness 
evaluations obtained across all data realizations and all sources. 
\begin{table}
\caption{\label{tab:temp_num} Statistical summary of the number of fitness evaluations
for the two hypotheses under which data realizations were generated. The 
sample minimum, maximum, average, and standard deviation (SD) of the number of fitness
evaluations are calculated over all the data realizations used for each hypothesis.}
\begin{center}
\begin{tabular}{|c|c|c|c|c|}
\hline
  & Min & Max  & Mean  & SD \\ \hline
 Signal Absent ($H_0$) & 163908 & 230808 & 210920 & 10758 \\
 Signal Present ($H_1$) & 193620 & 232716 & 221875 & 5373  \\
\hline
\end{tabular}
\end{center}
\end{table}

We see that the  mean number of fitness evaluations is slightly lower 
in the case where a signal is absent. Thus, particles have an enhanced 
tendency to exit the search space boundary when a signal is absent.
This may be a result of the
fact that the contrast between the values at the local maxima of the 
fitness function is less pronounced in this case. Since, most of the 
data from a GW detector consists of only noise, the fitness evaluation 
count for the signal absent case is more representative of the computational cost
that will be incurred in practice.

\section{Conclusion}
\label{Conclusion}
This paper presents a study of a PSO based approach to 
solving the computational challenge, stemming from the necessity to carry
out a high dimensional numerical optimization task, 
in a fully-coherent all-sky search for CBC signals.

At an astrophysically realistic signal strength {(e.g.,  
the ${\rm SNR}_{\rm opt}$ used here matches ${\rm SNR}_{\rm opt}=13$ for GW151226),}
we find that the best fitness value returned by 
PSO can approximate the GLRT quite effectively, suffering $\leq 2.5\%$ loss in 
detection probability,
while using $< 1/10$ the number of likelihood evaluations needed for the 
grid-based or Bayesian searches. 
It is important to emphasize here that we are not altering the standard representation 
of CBC waveforms or approximating the likelihood function in any way. Any alternative
scheme for likelihood or waveform calculation can be substituted without affecting the 
PSO algorithm itself.

A comparsion of our parameter estimation results with 
theoretical bounds derived from a Fisher information analysis is not meaningful
at the {SNR} value considered in this paper. 
This is because several studies have
shown~\cite{Balasubramanian_96,Vallisneri_07} that
these bounds are reached only at significantly
higher {${\rm SNR}_{\rm opt}$} values.


A direct comparison with existing parameter estimation 
results from Bayesian approaches is difficult, since
the definition of error in a Bayesian analysis differs from the Frequentist one. 
Error in a Bayesian analysis is  
a measure of the spread of the posterior probability distribution.
The latter can be obtained even for a single data realization. The Frequentist
error is a measure of the spread of the point estimates over an ensemble of 
data realizations. Nonetheless, pending a future apples-to-apples comparison of
Bayesian and Frequentist errors on identical data realizations, we find
that the best case error of $\sim 10$~${\rm deg}^2$
in sky position from PSO is near the expected 
ballpark at the value of ${\rm SNR}_{\rm opt}$ used here.
{For example, in~\cite{Rover_08} the sky location error for a signal with ${\rm SNR}_{\rm opt}=29.6, m_1=1.5M_\odot, m_2=2.0M_\odot$, and the HLV network is found to be $3\deg^2$.}

 Although our results have been obtained for the ideal case of Gaussian, stationary noise, 
the computational cost will not change significantly for real detector noise. Recall that
we are using PSO for only locating the global maximum of a fitness function. As long as
the nature of this fitness function, in terms of the density of local peaks and the 
contrast in their values, does not change drastically, PSO will have the same performance.
This is already evident when the computational cost of PSO is compared for the signal 
present and absent cases. We expect the 
change in the nature of the fitness function between these
two cases to be far more significant than that between ideal and real detector noise.

We have run PSO on a rectangular search region consisting of independent upper and lower bounds on each parameter.  This implies that the search range for each chirp time includes unphysical values. However, our results demonstrate that the probability of the global maximum straying into the unphysical region at the value of ${\rm SNR}_{\rm opt}$ used is negligible. This need not be true when the data contains pure noise, and this may affect detection performance by increasing the false alarm probability somewhat. A rigorous study of this issue is postponed to future work.

Our results show that PSO offers a promising approach to realize a constantly on,
fully-coherent all-sky CBC search. Future investigations should address the 
following outstanding issues. (i) A determination of wall-clock time savings 
after incorporating state-of-the-art waveform generation and likelihood evaluation
techniques. (ii) Reducing the instances of failure in locating the global maximum by 
trying out well-studied variants of PSO. For example, in~\cite{pso_pta}, 
the neighborhood best, rather than global best, variant of PSO was found to perform significantly
better. (iii) Extension of the analysis to the computationally more demanding case of aLIGO 
noise power spectral density.  


\section*{Acknowledgements}
We thank Y. Wang for sharing the KDE estimation codes for sky localization studies. Most of the computation was done on ``Thumper", a computer cluster funded by 
NSF MRI award CNS-1040430. T.S.W. acknowledges help from R.~Jackson, manager of Thumper, in using the cluster. We thank M.~Zanolin, E.~Schlegel, J.~Romano and S.~Mukherjee for helpful discussions. The contribution of S.D.M. to this paper was supported by NSF 
awards PHY-1505861 and HRD-0734800. 

\appendix
\section{Functional forms of phase parameters}
\label{app_A}


Let M, $\mu$ and $\eta$ denote the total mass, the reduced mass and the symmetric mass ratio of the compact binary system respectively. Let $f_*$ be the lower cutoff frequency of the detector. Then for $m_1 > m_2$, the chirp times are given by,


\begin{eqnarray}
\tau_{0} &=& \frac{5}{256\pi} f_{*}^{-1} (\frac{GM}{c^3}\pi  f_{*})^{-5/3}\eta^{-1},\\
\tau_{1} &=& \frac{5}{192\pi} f_{*}^{-1} (\frac{GM}{c^3}\pi f_{*})^{-1}\eta^{-1}\big(\frac{743}{336} + \frac{11}{4}\eta\big), \\
\tau_{1.5} &=& \frac{1}{8}f_{*}^{-1}(\frac{GM}{c^3}\pi  f_{*})^{-2/3}\eta^{-1}, \\
\tau_{2} &=&\frac{5}{128\pi}f_{*}^{-1}(\frac{GM}{c^3}\pi  f_{*})^{-1/3}\nonumber\\
&& \eta^{-1}\big(\frac{3058673}{1016064}+\frac{5429}{1008}\eta+\frac{617}{144}\eta^2 \big),
\end{eqnarray}
where
\begin{eqnarray}
M &=& (m_1 + m_2 ),\;\mu = \frac{m_1m_2}{M},\; \rm {and} \quad\eta =\frac{\mu}{M}.
\end{eqnarray}
All the four chirp time parameters are functions of $m_1$ and $m_2$, implying that only two of them are independent. We chose $\tau_0$ and $\tau_{1.5}$ as independent parameters to characterize the signal. Estimated values of $\tau_0$ and $\tau_{1.5}$ were used to derive values for $M$ and $\mu$ using the following equations.
\begin{eqnarray}
\mu&=&\frac{1}{16f_*^2}\bigg(\frac{5}{4\pi^4\tau_0\tau_{1.5}^2}\bigg)^{1/3}\bigg(\frac{G}{c^3}\bigg)^{-1}.\\ 
M &=& \frac{5}{32f_*}\bigg(\frac{\tau_{1.5}}{\pi^2\tau_0}\bigg)\bigg(\frac{G}{c^3}\bigg)^{-1}. \\ \nonumber
\end{eqnarray}   
The parameters $\alpha_i$, $i = 0, 1,\ldots,4$ in the phase function $\Psi(f)$ (Eq.~\ref{2pn_phase}) are given by,
\begin{eqnarray}
\alpha_0 &=& 2\pi f_* \frac{3\tau_0}{5},   \quad
\alpha_1 = 0 , \quad 
\alpha_2 = 2\pi f_*\tau_1,  \\
\alpha_3 &=& \textcolor{NavyBlue}{-}2\pi f_* \frac{3\tau_{1.5}}{2}, \nonumber \quad
\alpha_4 = 2\pi f_* 3\tau_2.
\end{eqnarray}

\end{document}